\pdfoutput=1
\documentclass[manuscript]{acmart}

\setcopyright{none}
\acmJournal{TOSEM}
\renewcommand\footnotetextcopyrightpermission[1]{}

\usepackage{graphicx}
\usepackage{booktabs}
\usepackage{array}
\usepackage{multirow}
\usepackage{tikz}
\usetikzlibrary{arrows.meta,positioning,shapes.geometric,fit,calc}
\usepackage{xcolor}

\newtheorem{question}{\textbf{RQ}}
\newcounter{myrq}
\begin{document}
	
	\title{PLC-BinX: A Cross-Platform Binary Code Analysis Framework for PLC Binaries}
	
	\author{Ang Jia}
	\affiliation{%
		\institution{Dalian University of Technology}
		\department{School of Software}
		\city{Dalian}
		\country{China}}
	
	\author{Yaxin Duan}
	\affiliation{%
		\institution{Dalian University of Technology}
		\department{School of Software}
		\city{Dalian}
		\country{China}}
	
	\author{He Jiang}
	\authornote{Corresponding author.}
	\email{jianghe@dlut.edu.cn}
	\affiliation{%
		\institution{Dalian University of Technology}
		\department{School of Software}
		\city{Dalian}
		\country{China}}
	
	\author{Zhenzhou Tian}
	\affiliation{%
		\institution{Xi'an University of Posts and Telecommunications}
		\department{School of Computer Science and Technology}
		\city{Xi'an}
		\country{China}}
	
	\author{Zhilei Ren}
	\affiliation{%
		\institution{Dalian University of Technology}
		\department{School of Software}
		\city{Dalian}
		\country{China}}
	
	\author{Xiaochen Li}
	\affiliation{%
		\institution{Dalian University of Technology}
		\department{School of Software}
		\city{Dalian}
		\country{China}}
	
	\begin{abstract}
	As emerging attacks increasingly target Industrial Control Systems (ICS), the security of Programmable Logic Controllers (PLCs) has become a critical concern. Binary Code Analysis (BCA), which enables analysts to analyze compiled programs, is essential for ICS security tasks such as deployed-binary auditing. However, automated BCA for PLC binaries remains challenging due to three key issues: heterogeneous binary formats across PLC platforms, entangled program semantics with runtime code, and limited semantic representations for downstream tasks. To resolve these challenges, we present PLC-BinX, a cross-platform BCA framework for PLC binaries. PLC-BinX applies a three-stage PLC binary analysis workflow, including cross-platform reverse engineering, core function identification, and function-level semantic representation, to analyze PLC binaries from four platforms: CODESYS v3, GEB, OpenPLC v2, and OpenPLC v3. We evaluate PLC-BinX on PLC-BEAD, which contains 2,431 PLC binaries across four platforms, using two downstream tasks: toolchain prediction and functionality prediction. Experimental results show that PLC-BinX achieves 100.00\% precision, recall, and F1 in toolchain prediction, and 51.43\% precision, 49.38\% recall, and 49.18\% F1 in functionality prediction over 22 labels.  These results demonstrate that PLC-BinX can transform raw PLC binaries into effective function-level semantic representations for PLC binary code analysis.
	\end{abstract}
	
	\keywords{PLC binary code analysis, cross-platform reverse engineering, function-level semantic representation}

	\ccsdesc[500]{Software and its engineering~Software reverse engineering}
	
	\maketitle
	
	\section{Introduction}
	
	Industrial Control Systems (ICS) are increasingly exposed to cyber threats that can affect not only digital assets but also physical processes. At the center of many ICS environments are Programmable Logic Controllers (PLCs), which directly execute control logic over sensors, actuators, and physical processes. For example, Stuxnet targeted Siemens industrial control systems and modified PLC logic to manipulate centrifuge operations while hiding abnormal behavior from operators~\cite{stuxnet}. Other studies have shown that vulnerabilities in widely deployed PLC runtime components can be abused to implant backdoors or gain control of industrial applications~\cite{codesys_runtime_backdoor}. These examples show that, after an attack, the logic executed by a PLC may no longer be the intended control program, but a modified or malicious version deployed on the controller.
	
	This observation makes deployed PLC binaries important evidence for security analysis. In post-attack forensics, incident response, and supply-chain auditing, analysts often need to inspect the binary running on the controller to determine what logic was executed and whether the control logic was modified. However, attackers rarely leave behind the source code, engineering project, or build configuration of their malicious logic. In many cases, the deployed binary is the only available artifact that reflects the actual PLC behavior. Therefore, Binary Code Analysis (BCA) is necessary for analyzing deployed PLC binaries and recovering higher-level program information from compiled PLC binaries.

	Existing studies have advanced the BCA of PLC binaries. For example, PLC-BEAD~\cite{plcbead} provides a benchmark of PLC binaries compiled from multiple platforms and programming languages, enabling systematic evaluation of BCA for PLC binaries. It further proposes PLCEmbed for functionality prediction and toolchain prediction from raw binary bytes. ICSREF~\cite{icsref} automates the reverse engineering of CODESYS v2-compiled PLC binaries and demonstrates how such capability can support both process-aware attack generation and defensive analysis. These efforts have advanced PLC binary analysis for security applications. However, existing methods still face \textbf{three challenges}, especially when moving to cross-platform PLC binary code analysis.
	
	\textbf{C1: Heterogeneous Binary Formats.}
	PLC binaries are highly heterogeneous across platforms. Existing reverse-engineering frameworks such as ICSREF can analyze CODESYS v2 binaries, but PLC binaries produced by different platforms expose substantially different binary structures. For example, CODESYS v3 applications are packaged as \texttt{.app} containers rather than standard executable files, GEB binaries appear as ARM ELF executables with generated POU-style symbols, and OpenPLC v2 and OpenPLC v3 binaries use \texttt{.exe} filename suffixes. These binaries differ in binary format, binary organization and runtime environment. Therefore, a method designed for one PLC platform cannot be directly applied to others.
	
	\textbf{C2: Entangled Program Semantics.}
	A PLC binary contains control logic, runtime and library functions, compiler-generated support functions, and platform-support functions. Only the control logic directly reflects user-defined PLC control semantics, while the other parts may dominate binary representations and obscure control-logic semantics. For example, PLCEmbed uses the entire binary for functionality prediction, where irrelevant semantics may dominate the learned representation and obscure the analysis of actual control logic.
	
	\textbf{C3: Limited Semantic Representation.}
	Existing PLC BCA methods provide limited support for higher-level semantic analysis. For example, PLCEmbed predicts functionality from raw byte sequences, but raw bytes do not explicitly expose program structures such as functions, basic blocks, or calls. Reverse-engineering frameworks such as ICSREF can recover useful information from CODESYS v2 binaries, but they do not provide representations for downstream tasks. Therefore, existing representations remain limited for systematic PLC binary analysis.

	To tackle the above challenges, this paper presents PLC-BinX, a cross-platform BCA framework for PLC binaries from CODESYS v3, GEB, OpenPLC v2, and OpenPLC v3. To address \textbf{C1: Heterogeneous Binary Formats}, PLC-BinX proposes a cross-platform reverse-engineering framework that disassembles heterogeneous PLC binaries to recover function identities, instruction-level semantics, control-flow structures, and function-call relations. To address \textbf{C2: Entangled Program Semantics}, PLC-BinX identifies core functions using platform-specific function-identification rules, while preserving runtime functions separately for provenance analysis. To address \textbf{C3: Limited Semantic Representation}, PLC-BinX constructs function-level semantic representations, including normalized instruction sequences and ACFGs (Attributed Control Flow Graphs), to capture instruction semantics and intra-function control-flow structure.
	
	We evaluate PLC-BinX on PLC-BEAD, which contains 2,431 PLC binaries across four platforms, using two downstream tasks: toolchain prediction and functionality prediction. Experimental results show that PLC-BinX achieves 100.00\% precision, recall, and F1 in toolchain prediction, and 51.43\% precision, 49.38\% recall, and 49.18\% F1 in functionality prediction over 22 labels. These results demonstrate that PLC-BinX can transform raw PLC binaries into effective function-level semantic representations for PLC binary code analysis.
	
	This paper makes the following contributions:
	
	\begin{itemize}
		\item To the best of our knowledge, PLC-BinX is the first framework that disassembles CODESYS v3, GEB, OpenPLC v2, and OpenPLC v3 binaries into function-level semantic representations for cross-platform downstream PLC binary analysis.
		
		\item We design a three-stage PLC binary analysis workflow in PLC-BinX, including cross-platform reverse engineering, core function identification, and function-level semantic representation.
		
		\item We evaluate PLC-BinX on PLC-BEAD, which contains 2,431 PLC binaries across four platforms, using two downstream tasks: toolchain prediction and functionality prediction. Experimental results show that PLC-BinX achieves 100.00\% precision, recall, and F1 in toolchain prediction, and 51.43\% precision, 49.38\% recall, and 49.18\% F1 in functionality prediction over 22 labels.
	\end{itemize}
	
	\section{Background}
	
	\subsection{PLCs in Industrial Control Systems}
	
	Industrial Control Systems (ICS) are used to monitor and control physical processes in domains such as manufacturing, energy, transportation, and water treatment. A typical ICS contains multiple layers, including supervisory software, communication networks, industrial controllers, and field devices. Among these components, PLCs play a central role because they directly bridge cyber decisions and physical process execution.
	
	Figure~\ref{fig:ics-plc-role} shows how PLCs connect supervisory components, field devices, and physical processes in a typical ICS. PLCs receive sensor inputs from the field, execute control logic, and output commands to actuators such as motors, valves, pumps, and relays. In a typical deployment, supervisory components such as engineering workstations, Human--Machine Interfaces (HMIs), or SCADA servers configure and monitor the control system, while PLCs execute time-critical control tasks close to the physical process. Because PLC behavior directly affects plant operation, the correctness and security of PLC software are critical to industrial safety and reliability.
	
	Most PLCs operate in a cyclic scan model. In each scan cycle, the controller first reads process inputs, then executes the user-defined control program, and finally writes outputs to the field devices. As a result, even small changes in PLC logic may directly alter the behavior of physical processes. This tight coupling between software and process control is what makes PLC software security particularly important in ICS.
	
	\subsection{PLC Source Code and Compilation Process}
	
	PLC applications are commonly developed in IEC 61131-3 languages, including Structured Text (ST), Ladder Diagram (LD), Function Block Diagram (FBD), Instruction List (IL), and Sequential Function Chart (SFC)~\cite{plcopen61131}. These languages provide different programming abstractions for industrial control tasks. For example, ST is a textual language similar to Pascal, LD is a graphical language derived from relay logic, and FBD expresses programs through interconnected function blocks.
	
\begin{figure}[t]
		\centering
		\begin{minipage}[b]{0.47\textwidth}
			\centering
			\includegraphics[width=0.98\linewidth]{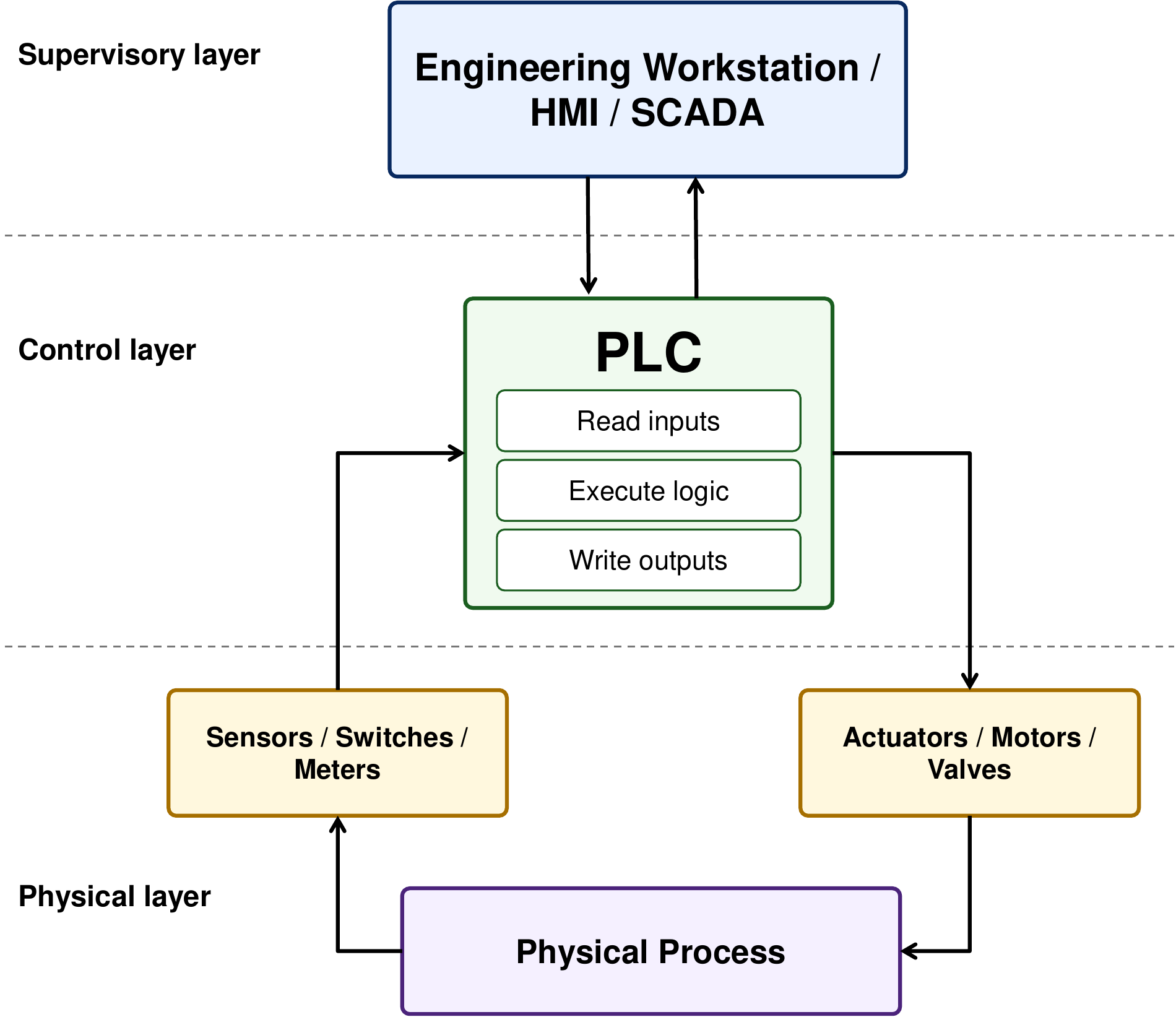}
			\Description{A typical industrial control system where PLCs connect supervisory components with sensors, actuators, and physical processes.}
			\vspace{15pt}
			\caption{The role of PLCs in a typical ICS.}
			\label{fig:ics-plc-role}
		\end{minipage}
		\hfill
		\begin{minipage}[b]{0.47\textwidth}
			\centering
			\includegraphics[width=0.95\linewidth]{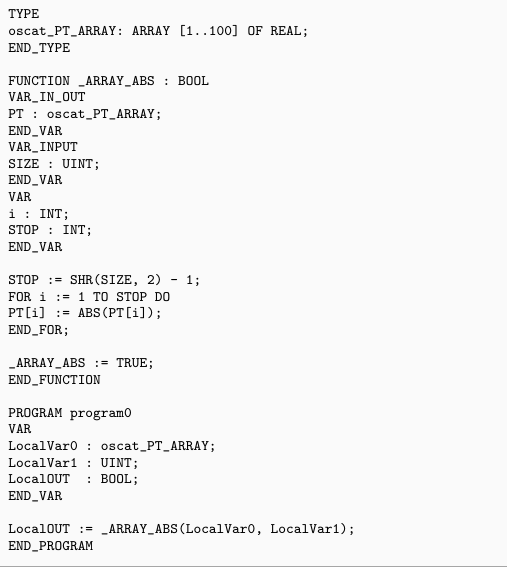}
			\Description{A Structured Text example showing a PLC program organization with a program, a user-defined function, variables, and a function call.}
			\caption{Example of PLC source organization using POUs.}
			\label{fig:pou-example}
		\end{minipage}
	\end{figure}

	A PLC project is usually organized into \textbf{Program Organization Units (POUs)}, such as \emph{programs}, \emph{functions}, and \emph{function blocks}. Programs typically define the main control logic, functions encapsulate reusable computations, and function blocks combine logic with internal state. As shown in Figure~\ref{fig:pou-example}, the program \texttt{program0} invokes the user-defined function \texttt{\_ARRAY\_ABS}, which iterates over an array and replaces each element with its absolute value. The example also shows typical PLC project elements, including data declarations and function interfaces. Compared with ordinary desktop software, PLC source projects contain not only algorithmic logic but also engineering context that binds the software to the physical process.
	
	After development, the source project is compiled by a platform-specific engineering environment and compiler toolchain into a deployable binary for the target PLC. For example, in CODESYS, the development system generates PLC-executable application code from the source project before downloading the application to the controller; this process checks allocations, data types, and library availability, and allocates memory addresses during code generation~\cite{codesys_generate_code}. A boot application can then be created as an \texttt{.app} file for execution on the PLC~\cite{codesys_boot_application}. In OpenPLC toolchains, IEC 61131-3 programs can be translated into lower-level C code by compilers such as \texttt{matiec} before the generated code is compiled into a binary~\cite{matiec}. Consequently, the final deployed binary contains both user-defined control logic and system-generated code introduced by the compiler, libraries, and runtime environment.

	\subsection{PLC Binary Formats}
	
	This paper analyzes the compiled PLC binary rather than the source project. Compared with the source-level view, the binary-level view is much less transparent: source-level names, POU boundaries, variable declarations, and engineering context may be transformed, partly or completely hidden depending on the target platform and compilation process. As a result, analysts cannot assume that the structure visible in the source project can be directly observed from the deployed binary.
	
	PLC binaries also vary substantially across platforms. Some platforms generate conventional executable files, while others package code and metadata into vendor-specific containers. Moreover, the deployed binary usually contains not only control logic, but also runtime and library functions, compiler-generated support functions, and platform-support functions. Therefore, a PLC binary is a composite compiled output shaped by the source project, the compiler toolchain, and the runtime environment, rather than a simple translation of the user program.

	From a security perspective, the deployed binary is often the only available form of the program for incident response, malware analysis, and supply-chain auditing.
	After an attack, analysts usually need to inspect the binary deployed on the controller. This makes PLC binary analysis a necessary step toward analyzing real-world industrial software.

	\section{Approach}

\begin{figure*}[t]
		\centering
		\includegraphics[width=0.92\textwidth]{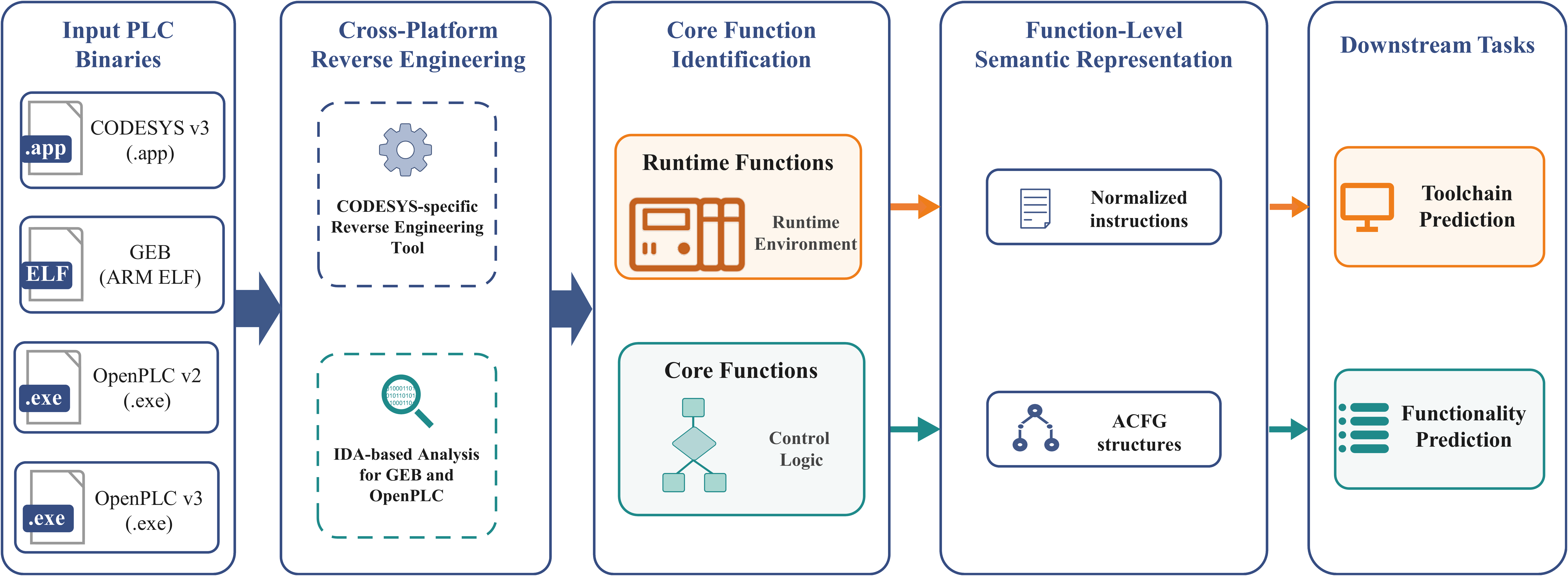}
	\Description{Overview of the PLC-BinX workflow from heterogeneous PLC binaries to cross-platform reverse engineering, core function identification, function-level semantic representation, and downstream tasks.}
	\caption{Overview of the PLC-BinX workflow.}
		\label{fig:overview}
	\end{figure*}

	The goal of PLC-BinX is to support effective PLC binary analysis for heterogeneous PLC binaries. As shown in Figure~\ref{fig:overview}, PLC-BinX takes PLC binaries from four platforms, including CODESYS v3, GEB, OpenPLC v2 and OpenPLC v3, as input. First, PLC-BinX performs cross-platform reverse engineering to recover function identities, instruction-level semantics, control-flow structures, and function-call relations from heterogeneous binary formats. Second, PLC-BinX performs core function identification to select functions that are more likely to represent PLC control logic, while preserving runtime functions separately for provenance analysis. Third, PLC-BinX constructs function-level semantic representations, including normalized instruction sequences and ACFGs, for recovered functions. To validate the usefulness of these function-level semantic representations, we use two downstream tasks: toolchain prediction and functionality prediction. Toolchain prediction validates whether the recovered functions can capture compiler and runtime provenance, while functionality prediction validates whether they can capture control behavior. The following subsections describe these stages in detail.

	\subsection{Input PLC Binaries}
	
	We use PLC-BEAD~\cite{plcbead} as the dataset for our study. To the best of our knowledge, PLC-BEAD is the only publicly available cross-platform PLC binary dataset for systematic PLC binary analysis. Table~\ref{tab:plc-binary-forms} summarizes the PLC platforms, binary forms, and dataset scale of this dataset. The dataset contains 729 PLC source programs and 2,431 compiled binaries across four platform settings. Not every source program has a corresponding binary on every platform, because some programs cannot be compiled under specific toolchains.
	
	The four platform settings cover vendor-specific application containers and executable files. CODESYS v3 applications are distributed as vendor-specific \texttt{.app} containers. GEB binaries are ARM ELF executables, while OpenPLC v2 and OpenPLC v3 binaries use \texttt{.exe} filename suffixes. These differences show why PLC binary analysis requires cross-platform reverse engineering instead of a single uniform parsing strategy.
	
	\begin{table*}[t]
		\centering
		\footnotesize
		\begin{minipage}[t]{0.48\textwidth}
			\centering
			\caption{PLC platforms and binary forms in the dataset.}
			\label{tab:plc-binary-forms}
			\begin{tabular}{@{}llr@{}}
				\toprule
				Platform & Binary form & \# Binaries \\
				\midrule
				CODESYS v3~\cite{codesys_boot_application} & \texttt{.app} container & 555 \\
				GEB~\cite{plcbead} & ARM ELF executable & 617 \\
				OpenPLC v2~\cite{openplc} & \texttt{.exe} file & 619 \\
				OpenPLC v3~\cite{openplc} & \texttt{.exe} file & 640 \\
				\midrule
				Total & -- & 2,431 \\
				\bottomrule
			\end{tabular}
		\end{minipage}
		\hfill
		\begin{minipage}[t]{0.48\textwidth}
			\centering
			\caption{Core-function naming patterns across PLC platforms.}
			\label{tab:core-platforms}
			\begin{tabular}{@{}>{\raggedright\arraybackslash}m{0.27\linewidth}>{\raggedright\arraybackslash}m{0.62\linewidth}@{}}
				\toprule
				Platform & Core-function naming patterns \\
				\midrule
				CODESYS v3 & \texttt{PLC\_PRG}, metadata-visible POU/function/method symbols \\
				GEB & \texttt{dt\_PR\_*}, \texttt{dt\_FN\_*}, \texttt{dt\_FB\_*} \\
				OpenPLC v2 and v3 & \texttt{PROGRAM0\_body\_\_}, \texttt{\_\_PROGRAM0\_\_*} \\
				\bottomrule
			\end{tabular}
		\end{minipage}
	\end{table*}
	
	\subsection{Cross-Platform Reverse Engineering}
	
	Because PLC platforms use different binary formats, PLC-BinX applies platform-specific reverse-engineering procedures. For standard executable formats, including GEB ARM ELF binaries and OpenPLC v2 and OpenPLC v3 executable files with \texttt{.exe} filename suffixes, PLC-BinX uses IDA-based analysis for GEB and OpenPLC. For CODESYS v3, whose \texttt{.app} files are vendor-specific application containers rather than standard executables, PLC-BinX implements a CODESYS-specific reverse-engineering tool. Although these procedures differ internally, their outputs are normalized into recovered function-level information, including function identities, function boundaries, assembly instructions, control-flow structures, and function-call relations.
	
	\subsubsection{IDA-based analysis for GEB and OpenPLC}
	
	For GEB and OpenPLC binaries, PLC-BinX first loads the executable into IDA and performs batch disassembly. IDA recovers function names, function boundaries, assembly instructions, basic blocks, control-flow edges, and direct call relationships. These outputs provide function identities, instruction-level semantics, control-flow structures, and function-call relations for later analysis. Function boundaries and assembly instructions define the instruction sequence of each recovered function, basic blocks and control-flow edges support CFG and ACFG construction, and direct call relationships support call-related representations such as FCGs (Function Call Graphs). The recovered names also support later core function identification. 
	
	\subsubsection{CODESYS-specific reverse-engineering tool}
	
	CODESYS v3 requires a separate reverse-engineering method because its \texttt{.app} files are vendor-specific application containers rather than standard executable files. As a result, they cannot be directly analyzed by the ordinary IDA-based batch disassembly process, which assumes a conventional executable loader, section layout, entry information, and symbol information. PLC-BinX therefore implements a CODESYS-specific reverse-engineering tool with three main steps: executable code discovery, function recovery, and symbol recovery. The recovered CODESYS function-level information is then converted into the function-level information consistent with the IDA-based platforms.

	\textbf{Executable code discovery.}
	Since a CODESYS v3 \texttt{.app} file does not provide a standard executable header or section table, PLC-BinX treats the file as a binary container and locates executable code by searching for function-level boundaries. In the studied CODESYS v3 \texttt{.app} artifacts, the executable payload is decoded as ARM32 little-endian code. PLC-BinX scans 4-byte aligned offsets for ARM32 function-entry patterns. In our implementation, an entry candidate is identified from common stack-frame setup patterns, such as \texttt{STMDB SP!, \{...\}} or its assembler alias \texttt{PUSH \{...\}}, followed by frame-register setup instructions such as \texttt{MOV FP, SP}, \texttt{MOV SL, SP}, or \texttt{MOV IP, SP}. Starting from each candidate entry, PLC-BinX scans forward to return-like ARM instructions, such as \texttt{LDMFD SP!, \{..., PC\}}, \texttt{POP \{..., PC\}}, or \texttt{MOV PC, LR}, to estimate the end of the function slice.

	\textbf{Function recovery.}
	After discovering executable slices, PLC-BinX uses Capstone~\cite{capstone} to disassemble each slice as an ARM32 little-endian instruction stream. Each retained slice is treated as a recovered function interval and assigned an address-based identity, such as \texttt{sub\_000145E0}. Because compiler optimizations may inline helper routines or transform tail calls, these intervals are binary-level function slices rather than guaranteed one-to-one source-level functions.
	
	PLC-BinX then constructs function-level structure from the decoded instructions. It identifies basic block leaders from the function entry, decoded branch targets, and fall-through positions after conditional branches. It further recovers control-flow edges from branch instructions and sequential fall-through relations. Direct calls are recovered from call instructions such as \texttt{BL} and \texttt{BLX}, while indirect transfers and return-like instructions are recorded separately. The resulting recovered function contains assembly instructions, basic blocks, control-flow edges, and function-call relations, which are exported in an IDA-compatible format for downstream function-level semantic representation.

	\textbf{Symbol recovery.}
	Finally, PLC-BinX parses metadata records embedded in the \texttt{.app} file to recover binary-visible symbols. These records may contain symbol names and associated code pointers. PLC-BinX uses these code pointers to associate metadata-visible symbols with recovered functions. In this way, metadata-visible units such as \texttt{PLC\_PRG} and user-defined POU functions receive recovered names, while runtime helpers and support functions without metadata-visible names remain represented by address-based identities. For example, in \texttt{\_ARRAY\_ABS.app}, PLC-BinX associates metadata symbols such as \texttt{PLC\_PRG} and \texttt{\_ARRAY\_ABS} with recovered functions, producing named recovered functions for later core function identification and function-level semantic representation.

	\subsection{Core Function Identification}
	
	Reverse engineering is a necessary step for PLC BCA, but it does not distinguish control logic semantics from platform runtime behavior.
	During the development of PLC applications, the developers only write the control logic, while the control logic is compiled and aggregated with runtime code to generate the final binaries. As a consequence, PLC binaries contain not only control logic functions, but also runtime and library functions, compiler-generated support functions, and platform-support functions. These non-core functions may obscure the semantics of binaries, making the real control logic hidden in the enormous binary functions. PLC-BinX therefore identifies core functions that represent the control logic and then separates recovered functions into two categories:
	
	\begin{itemize}
		\item \textbf{Core functions:} functions most likely to implement PLC control-logic semantics. These functions describe what the PLC program does and therefore represent the core functionality of a PLC binary.
		
		\item \textbf{Runtime functions:} non-core functions outside the control logic, including runtime and library functions, compiler-generated support functions, and platform-support functions. Their instruction sequences, function hashes, and call patterns do not represent the core control logic, but contain the toolchain-related information.
	\end{itemize}

	Table~\ref{tab:core-platforms} shows the platform-specific naming patterns used for core function identification. For CODESYS v3, PLC-BinX uses symbols recovered from \texttt{.app} records to identify core functions. The \texttt{PLC\_PRG} symbol is treated as the primary control logic function because it represents the main PLC program organization unit. Following the sub-FCG of \texttt{PLC\_PRG}, the recovered functions in the call-chain are also included as core functions. 
	
	For GEB, the main program body is usually rooted at \texttt{dt\_PR\_program0\_exec}, while user-defined functions and function blocks appear as \texttt{dt\_FN\_*} and \texttt{dt\_FB\_*}. PLC-BinX selects \texttt{dt\_PR\_program0\_exec} as the primary control logic function and follows recovered calls to collect functions that match these generated control-logic naming patterns. When this traversal reaches GEB runtime functions, PLC-BinX records the outgoing calls but does not expand the runtime functions into the core-function set.
	
	For OpenPLC v2 and OpenPLC v3, functions whose names match patterns \texttt{PROGRAM0\_body\_\_} are used as the primary control logic function because they implement the main program body, and generated control logic functions are identified by names such as \texttt{\_\_PROGRAM0\_\_*}. Starting from the entry, PLC-BinX follows recovered calls to collect candidate control logic functions and filters functions that match standard C and C++ library patterns, OpenPLC runtime patterns, or platform-support patterns. The same identification rule is used for OpenPLC v2 and v3 because their generated binaries follow a similar binary organization.
	
	After core function identification, PLC-BinX treats the other recovered functions as runtime functions. In this work, the term ``runtime functions'' is used broadly to include runtime functions, library functions, compiler-generated support functions, and platform-support functions. These functions do not represent the control logic of PLC, but they are preserved because they contain distinctive toolchain and runtime fingerprints.

	\subsection{Function-Level Semantic Representation}
	
	After core function identification, PLC-BinX constructs function-level semantic representations for the recovered binary functions. Both core functions and runtime functions can be represented as normalized instruction sequences or ACFGs. For downstream models, these function-level representations can be used directly or further transformed into task-specific features, such as runtime-function fingerprints.
	
	\textbf{Normalized instructions.}
	For each selected function, PLC-BinX normalizes assembly instructions before using them for downstream tasks. One important finding from our reverse-engineering process is that, although the studied binaries are produced by different PLC platforms, they all use the same ARM32 instruction set in the PLC-BEAD dataset. Therefore, we use the same normalization process for all these binaries. As shown in Table~\ref{tab:instruction-normalization}, the normalization keeps the opcode semantics while abstracting unstable operands. Mnemonics are lowercased, registers are replaced with \texttt{reg}, memory operands are represented by abstract memory tokens, and numeric immediates are replaced with \texttt{imm}. For example, \texttt{MOV R3, R0} becomes \texttt{mov reg reg}, while \texttt{CMP R3, \#2} becomes \texttt{cmp reg imm}. Direct calls are handled specially because they carry semantic information about inter-function dependencies. If the call target can be resolved to a recovered function name, PLC-BinX keeps a name-aware call token such as \texttt{call\_name:\_ARRAY\_ABS}; otherwise, unresolved direct calls and indirect transfers are represented using generic tokens such as \texttt{call\_addr} and \texttt{call\_indirect}. This representation preserves semantic information available in the recovered binary while reducing sensitivity to unstable low-level operands.
	
	\textbf{ACFG structures.}
	ACFG structures can further reveal the semantics of control logic in core functions. Each node corresponds to a basic block, and its primary feature is the normalized instruction sequence inside that basic block. The normalization follows Table~\ref{tab:instruction-normalization}. As shown in Table~\ref{tab:acfg-node-information}, PLC-BinX also adds other features, including recovered function or call-target names, lightweight block-level context, and function-category information as the node attributes when such information can be recovered. Edges are derived from decoded control-transfer instructions. A conditional branch such as \texttt{BNE loc\_1} adds both a target edge and a fall-through edge, while an unconditional branch such as \texttt{B loc\_2} or \texttt{JMP loc\_2} adds only a target edge. Return-like instructions such as \texttt{BX LR}, \texttt{RET}, or \texttt{MOV PC, LR} mark function exits.
	
	\begin{table*}[t]
		\centering
		\footnotesize
		\begin{minipage}[t]{0.39\textwidth}
			\centering
			\caption{Examples of instruction normalization.}
			\label{tab:instruction-normalization}
			\begin{tabular}{@{}ll@{}}
				\toprule
				Raw instruction & Normalized tokens \\
				\midrule
				\texttt{MOV R3, R0} & \texttt{mov reg reg} \\
				\texttt{CMP R3, \#2} & \texttt{cmp reg imm} \\
				\texttt{LDR R0, [SP,\#8]} & \texttt{ldr reg mem} \\
				\texttt{BL \_ARRAY\_ABS} & \texttt{bl call\_name:\_ARRAY\_ABS} \\
				\texttt{BL 0x4012A0} & \texttt{bl call\_addr} \\
				\texttt{MOV PC, R6} & \texttt{mov call\_indirect} \\
				\bottomrule
			\end{tabular}
		\end{minipage}
		\hfill
		\begin{minipage}[t]{0.57\textwidth}
			\centering
			\caption{Examples of ACFG node information.}
			\label{tab:acfg-node-information}
			\begin{tabular}{@{}p{0.30\linewidth}p{0.45\linewidth}@{}}
				\toprule
				Node information & Example values \\
				\midrule
				Instruction tokens & \texttt{mov reg reg}; \texttt{cmp reg imm} \\
				Recovered name tokens & \texttt{func:PLC\_PRG}; \texttt{call\_name:\_ABS} \\
				Instruction count & \texttt{bb\_len:9-16} \\
				Control-flow degree & \texttt{bb\_in:1}; \texttt{bb\_out:2} \\
				Function category & \texttt{core\_function} \\
				\bottomrule
			\end{tabular}
		\end{minipage}
	\end{table*}
	
	\subsection{Downstream Tasks}
	
	Based on the recovered function-level semantic representations of core and runtime functions, we use two downstream tasks, toolchain prediction and functionality prediction, to validate effectiveness of the recovered representations. Toolchain prediction detects which compiler and runtime produced the binary, so it validates whether recovered functions preserve provenance information. Functionality prediction detects what control behavior the PLC program implements, so it validates whether recovered functions preserve control-logic semantics.
	
\begin{figure*}[t]
		\centering
		\begin{minipage}[b]{0.47\textwidth}
			\centering
			\includegraphics[width=0.96\linewidth]{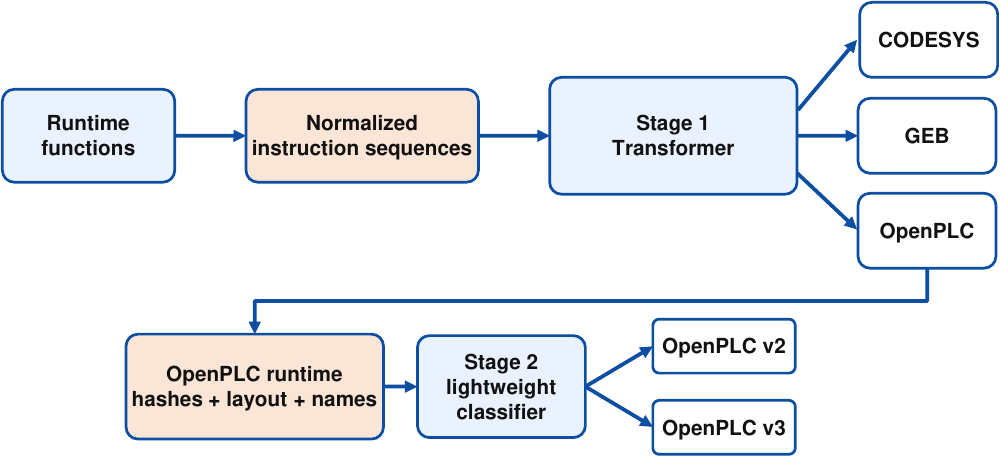}
			\Description{Two-stage toolchain prediction workflow using runtime-function representations and OpenPLC runtime fingerprints.}
			\caption{Workflow of PLC-BinX for toolchain prediction.}
			\label{fig:toolchain}
		\end{minipage}
		\hfill
		\begin{minipage}[b]{0.47\textwidth}
			\centering
			\includegraphics[width=0.96\linewidth]{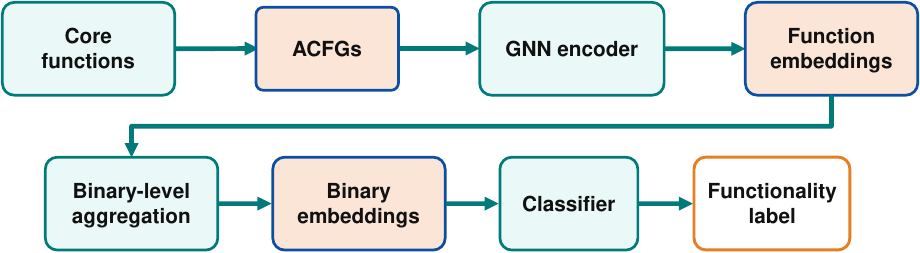}
			\Description{Hierarchical functionality prediction workflow using core-function ACFGs, function-level graph encoding, binary-level aggregation, and classification.}
			\caption{Workflow of PLC-BinX for functionality prediction.}
			\label{fig:functionality}
		\end{minipage}
	\end{figure*}

	\textbf{Toolchain prediction.} We take the runtime functions as input to toolchain prediction as they contain rich toolchain-related information.
	As shown in Figure~\ref{fig:toolchain}, we implement toolchain prediction as a two-stage downstream task. The first stage performs coarse-grained platform-family prediction, distinguishing CODESYS, GEB, and OpenPLC. The second stage is applied only to binaries predicted as OpenPLC and further distinguishes OpenPLC v2 from OpenPLC v3. This design is motivated by our observations. Across CODESYS, GEB, and OpenPLC, binaries differ substantially in file format, function organization, and runtime structure. Therefore, the coarse platform family can be easily identified from the normalized instruction sequence. In contrast, OpenPLC v2 and OpenPLC v3 are much closer: they compile the same ST programs into highly similar binaries, but slightly differ in the surrounding runtime functions and platform-support functions. The two-stage design therefore facilitates coarse-grained family classification and fine-grained version classification.
	
	In the first stage, runtime functions are represented as normalized instruction sequences. These sequences are then concatenated into a binary-level token sequence and fed into a Transformer classifier to predict one of three platform families: CODESYS, GEB, or OpenPLC. Runtime functions are used because they are introduced by the compiler, runtime, and platform environment, whereas core functions are mainly derived from user-written PLC logic. Therefore, normalized instruction sequences of runtime functions provide more stable and platform-specific features for toolchain prediction, while representations derived from core functions may vary with the control logic.

	In the second stage, PLC-BinX focuses on OpenPLC samples and derives runtime-function fingerprints to distinguish OpenPLC v2 from OpenPLC v3. These fingerprints are task-specific binary-level features, including hashes of normalized instruction sequences of runtime functions, layout features such as the number and size distribution of runtime functions, and recovered runtime function names. A lightweight classifier based on runtime-function fingerprints then predicts the final OpenPLC version label, either OpenPLC v2 or OpenPLC v3.

	\textbf{Functionality prediction.} We take the core functions as input to functionality prediction as they correspond to the control logic of the PLC.
	For functionality prediction, PLC-BinX adopts a hierarchical ACFG-GNN architecture over recovered core functions, as shown in Figure~\ref{fig:functionality}. 
	
	The model contains two levels of representation learning. At the function level, a GNN encoder propagates information along the ACFG edges and computes a graph-level embedding for each core function. This embedding summarizes the instruction semantics and control-flow structure inside the function. At the binary level, an aggregation layer combines the embeddings of all core functions into a single binary representation. The final classifier predicts the functionality label from this binary-level representation. This hierarchical design allows PLC-BinX to preserve intra-function control-flow semantics and aggregate them into the control logic semantics of a PLC binary.

	\section{Experiments}
	\label{sec:experiments}
	
	In this section, we first examine how recovered PLC binary functions are distributed across core functions and runtime functions. We then use toolchain prediction and functionality prediction as downstream tasks to evaluate how effective PLC-BinX is compared to existing works. We further conduct an ablation study to analyze the importance of task-specific function-type selection. In general, we answer the following research questions.
	
	\begin{question}
		\label{RQ-function-distribution}
		\textit{How are core functions and runtime functions distributed across PLC platforms?}
	\end{question}
	\addtocounter{myrq}{1}
	
	\begin{question}
		\label{RQ-toolchain}
		\textit{How effective is PLC-BinX when conducting toolchain prediction compared with PLCEmbed?}
	\end{question}
	\addtocounter{myrq}{1}
	
	\begin{question}
		\label{RQ-functionality}
		\textit{How effective is PLC-BinX when conducting functionality prediction compared with PLCEmbed?}
	\end{question}
	\addtocounter{myrq}{1}
	
	\begin{question}
		\label{RQ-ablation}
		\textit{How important is the task-specific function-type selection when conducting downstream tasks?}
	\end{question}
	\addtocounter{myrq}{1}
	
	\subsection{Study Setup}
	
	\subsubsection{Evaluation Dataset.}
	We use PLC-BEAD~\cite{plcbead} as the evaluation dataset. The dataset contains 2,431 compiled binaries. The stage-1 toolchain family prediction task uses all 2,431 binaries, including 555 CODESYS samples, 617 GEB samples, and 1,259 OpenPLC samples. The OpenPLC version stage uses 619 OpenPLC v2 samples and 640 OpenPLC v3 samples. The functionality prediction task uses 2,430 labeled binaries across 22 functionality labels, because one CODESYS v3 binary in PLC-BEAD does not have a functionality label.
	
	\subsubsection{Experimental Setting.}
	All inputs to PLC-BinX are PLC binaries. Source code is not used as an input to PLC-BinX or to train downstream models. It is used only for dataset labels and offline auditing of recovered CODESYS functions. All reported downstream-task experiments use ten-fold program-level cross-validation with an 8:1:1 partition. Specifically, PLC programs are divided into ten folds. In each run, eight folds are used for training, one fold is used for validation to select the model from epochs with the best performance, and one fold is used for testing using the selected model. Across the ten runs, each valid binary appears in a test fold once. The split is performed at the PLC-program level, so binaries compiled from the same PLC program are assigned to the same fold and will not be used for both training and testing. For the PLCEmbed baseline, we use the open-source GitHub implementation. The difference is the data partitioning: we evaluate PLCEmbed under the same ten-fold program-level partition used for PLC-BinX.
	
	\subsubsection{Evaluation Metrics.}
	For the downstream tasks, we report precision, recall, and F1. Unless otherwise specified, weighted averages are computed by the number of binaries in each label. We also report per-label precision, recall, and F1 so that the comparison between PLCEmbed and PLC-BinX can be inspected at each toolchain or functionality label. For the ablation study, we use the same ten-fold program-level partition and the same metrics as the corresponding downstream task.
	
	\subsubsection{Implementation.}
	In the cross-platform reverse engineering stage, PLC-BinX uses IDA~\cite{ida} to implement the reverse engineering process for GEB and OpenPLC binaries, and implements the proposed CODESYS-specific reverse-engineering tool to disassemble the binaries from CODESYS. In the toolchain prediction, the toolchain family classifier uses a Transformer encoder~\cite{vaswani2017attention} while the OpenPLC version classifier uses a lightweight linear classifier.
	In the functionality prediction, PLC-BinX implements a hierarchical ACFG-GNN. At the function level, each core-function ACFG is encoded by a GraphSAGE-style message-passing network~\cite{hamilton2017graphsage}. At the binary level, function embeddings from the same binary are aggregated with an attention-based pooling layer and max pooling, and the resulting binary embedding is fed into a two-layer feed-forward classifier.
	
	\subsection{RQ~\ref{RQ-function-distribution}: Core and Runtime Function Distribution}
	
	We first characterize how recovered functions are distributed after platform-specific core function identification. Table~\ref{tab:core-runtime-distribution} presents the distribution of core functions and runtime functions for each platform and measures them by two metrics: the quantity of recovered functions and the size of recovered functions.
	
	\begin{table*}[t]
		\caption{Distribution of core and runtime functions across PLC platforms.}
		\label{tab:core-runtime-distribution}
		\centering
		\footnotesize
		\begin{tabular*}{\textwidth}{@{\extracolsep{\fill}}lrrrrrrrrr@{}}
			\toprule
			\multirow{3}{*}{Platform} & \multirow{3}{*}{Binaries} & \multicolumn{4}{c}{Core functions} & \multicolumn{4}{c}{Runtime functions} \\
			\cmidrule(lr){3-6}\cmidrule(l){7-10}
			& & \multicolumn{2}{c}{Quantity} & \multicolumn{2}{c}{Size} & \multicolumn{2}{c}{Quantity} & \multicolumn{2}{c}{Size} \\
			\cmidrule(lr){3-4}\cmidrule(lr){5-6}\cmidrule(lr){7-8}\cmidrule(l){9-10}
			& & Total & Per binary & Mean & Median & Total & Per binary & Mean & Median \\
			\midrule
			CODESYS v3 & 555 & 1,184 & 2.13 & 2,930.6 & 604.0 & 219,901 & 396.22 & 359.2 & 116.0 \\
			GEB & 617 & 2,042 & 3.31 & 339.7 & 142.0 & 754,223 & 1,222.40 & 86.6 & 68.0 \\
			OpenPLC v2 & 619 & 1,816 & 2.93 & 517.8 & 260.0 & 594,850 & 960.99 & 165.4 & 48.0 \\
			OpenPLC v3 & 640 & 1,947 & 3.04 & 537.4 & 264.0 & 690,844 & 1,079.44 & 174.2 & 48.0 \\
			\bottomrule
		\end{tabular*}
	\end{table*}
	
	The quantity distribution shows that PLC binaries contain a small number of core functions. Across the four platforms, each binary contains only 2.13--3.31 core functions on average. This small number is consistent with the organization of PLC programs, where control logic is usually concentrated in the main program unit and a limited number of user-defined functions or function blocks. In contrast, runtime functions are much more numerous. GEB contains 754,223 runtime functions in total, corresponding to 1,222.40 per binary on average. OpenPLC v2 and OpenPLC v3 contain 960.99 and 1,079.44 runtime functions per binary, respectively. CODESYS v3 also contains a large surrounding runtime context, with 219,901 runtime functions and 396.22 runtime functions per binary on average.
	
	The size distribution further shows that the core functions and runtime functions have different volumes. CODESYS v3 core functions are much larger on average than those of the other platforms, and their mean size is also much larger than their median size. This indicates that a small number of large core functions dominate the average. Runtime functions are typically smaller than core functions: GEB runtime functions have an average size of 86.6 bytes, while its core functions have an average size of 339.7 bytes. OpenPLC v2 and v3 have similar core-function sizes, but OpenPLC v3 contains more runtime functions, consistent with the additional communication-support functions recovered from OpenPLC v3 binaries.

	\noindent\textbf{Answering RQ~\ref{RQ-function-distribution}:}
	Core functions are sparse across PLC binaries, with only 2.13--3.31 core functions per binary on average, while runtime functions are substantially more numerous and platform-dependent, ranging from 396.22 to 1,222.40 functions per binary. This distribution supports our task-specific design: runtime functions are excluded from functionality prediction because they can obscure the main control logic from the core functions, while runtime functions are used for toolchain prediction because they can provide fine-grained distinctions between different platforms.
	
	\subsection{RQ~\ref{RQ-toolchain}: Toolchain Prediction}
	
	We then evaluate how effective PLC-BinX is when conducting toolchain prediction compared with PLCEmbed. Table~\ref{tab:toolchain} reports the precision, recall, and F1 of PLC-BinX and PLCEmbed for all platforms. In general, PLCEmbed reaches 91.63\% precision, 91.62\% recall, and 91.62\% F1, while PLC-BinX reaches 100.00\% precision, recall, and F1.
	
	\begin{table*}[t]
		\caption{Toolchain prediction results under ten-fold program-level evaluation.}
		\label{tab:toolchain}
		\centering
		\footnotesize
		\begin{tabular*}{\textwidth}{@{\extracolsep{\fill}}llrrrrrr@{}}
			\toprule
			\multirow{2}{*}{Platform} & \multirow{2}{*}{Version} & \multicolumn{3}{c}{PLCEmbed} & \multicolumn{3}{c}{PLC-BinX} \\
			\cmidrule(lr){3-5}\cmidrule(l){6-8}
			& & Precision & Recall & F1 & Precision & Recall & F1 \\
			\midrule
			CODESYS  & v3 & 99.96\% & 99.90\% & 99.93\% & 100.00\% & 100.00\% & 100.00\% \\
			GEB & -- & 98.92\% & 98.81\% & 98.86\% & 100.00\% & 100.00\% & 100.00\% \\
			\multirow{2}{*}{OpenPLC} & v2 & 82.77\% & 83.55\% & 83.16\% & 100.00\% & 100.00\% & 100.00\% \\
			& v3 & 85.94\% & 85.33\% & 85.63\% & 100.00\% & 100.00\% & 100.00\% \\
			\midrule
			Avg. & -- & 91.63\% & 91.62\% & 91.62\% & 100.00\% & 100.00\% & 100.00\% \\
			\bottomrule
		\end{tabular*}
	\end{table*}
	
	The main difference between PLC-BinX and PLCEmbed is that PLC-BinX can better classify OpenPLC v2 and OpenPLC v3. This can be attributed to the two-stage architecture of PLC-BinX. The fine-grained distinctions between OpenPLC v2 and OpenPLC v3 mainly come from the runtime functions: OpenPLC v3 adds communication-related runtime functions associated with ENIP, PCCC, and Modbus, whereas OpenPLC v2 exposes WiringPi and hardware input and output functions such as \texttt{wiringPiSetup}, \texttt{digitalRead}, \texttt{digitalWrite}, and \texttt{pwmWrite}. 
	The recovered function semantic representations of PLC-BinX help distinguish these distinctions, while PLCEmbed only takes the raw binary as input, missing these important clues. As a result, PLC-BinX is more effective for toolchain prediction.
	
	\noindent\textbf{Answering RQ~\ref{RQ-toolchain}:}
	PLC-BinX reaches 100.00\% precision, recall, and F1 in toolchain prediction. Compared with existing works, PLC-BinX can identify fine-grained distinctions between different platforms.
	
	\subsection{RQ~\ref{RQ-functionality}: Functionality Prediction}
	
	This RQ evaluates how effective PLC-BinX is when conducting functionality prediction. Table~\ref{tab:function-per-label} reports the precision, recall, and F1, together with the number of test binaries for each label after combining the test-fold predictions. In general, PLC-BinX reaches 51.43\% precision, 49.38\% recall, and 49.18\% F1. In comparison, PLCEmbed only reaches 5.11\% precision, 11.85\% recall, and 4.90\% F1 under the same program-level split.

	Note that the PLCEmbed results reported here differ from those in the original PLC-BEAD paper because the training and test partitions are different. We re-evaluate PLCEmbed under the same ten-fold program-level partition
	used for PLC-BinX, where binaries compiled from the same  PLC program are assigned to the same fold. This prevents binaries compiled from the same source program from appearing in both training and test sets.

	PLC-BinX obtains high F1 results on labels such as \texttt{Network\_1} (71.93\%), \texttt{Vector\_Math} (67.53\%), \texttt{Time\_and\_Date} (63.55\%), \texttt{Complex\_Math} (60.39\%), and \texttt{Ctrl\_Mods} (59.12\%), showing that recovered core functions and their ACFG-based function-level semantic representations support PLC functionality prediction. The remaining difficult labels include \texttt{Arithmetic\_Func} (17.28\%), \texttt{Basic\_Other\_Func} (25.67\%), \texttt{Signal\_Proc} (32.35\%), and \texttt{Measure\_Mods} (33.33\%). These classes are still not fully captured by the current function-level semantic representation. We will discuss the failure modes of PLC-BinX in Section \ref{sec:failuremode}.
	
	\begin{table*}[t]
		\caption{Functionality prediction results under ten-fold program-level evaluation.}
		\label{tab:function-per-label}
		\centering
		\footnotesize
		\setlength{\tabcolsep}{2.5pt}
		\renewcommand{\arraystretch}{0.92}
		\begin{tabular*}{\textwidth}{@{\extracolsep{\fill}}p{0.20\textwidth}rrrrrrr@{}}
			\toprule
			\multirow{2}{*}{Functionality label} & \multirow{2}{*}{Number} & \multicolumn{3}{c}{PLCEmbed} & \multicolumn{3}{c}{PLC-BinX} \\
			\cmidrule(lr){3-5}\cmidrule(l){6-8}
			& & Precision & Recall & F1 & Precision & Recall & F1 \\
			\midrule
			Actuators\_HVAC & 80 & 0.00\% & 0.00\% & 0.00\% & 54.17\% & 32.50\% & 40.62\% \\
			Arithmetic\_Func & 44 & 0.00\% & 0.00\% & 0.00\% & 18.92\% & 15.91\% & 17.28\% \\
			Array\_Buffer\_Mem\_List & 115 & 0.00\% & 0.00\% & 0.00\% & 52.68\% & 51.30\% & 51.98\% \\
			Basic\_Dev\_Drivers & 78 & 0.00\% & 0.00\% & 0.00\% & 37.97\% & 38.46\% & 38.22\% \\
			Basic\_Other\_Func & 75 & 0.00\% & 0.00\% & 0.00\% & 21.43\% & 32.00\% & 25.67\% \\
			Building\_Other\_Func & 69 & 0.07\% & 4.35\% & 0.13\% & 59.65\% & 49.28\% & 53.97\% \\
			Calculations & 80 & 0.00\% & 0.00\% & 0.00\% & 38.76\% & 62.50\% & 47.85\% \\
			Complex\_Math & 104 & 0.00\% & 0.00\% & 0.00\% & 50.99\% & 74.04\% & 60.39\% \\
			Ctrl\_Mods & 127 & 0.00\% & 0.00\% & 0.00\% & 55.10\% & 63.78\% & 59.12\% \\
			Latches\_FlipFlop\_ShiftReg & 64 & 0.00\% & 0.00\% & 0.00\% & 56.76\% & 32.81\% & 41.58\% \\
			Logic\_Mods & 216 & 13.81\% & 28.70\% & 18.41\% & 53.39\% & 58.33\% & 55.75\% \\
			Mathematics & 274 & 4.34\% & 1.82\% & 2.39\% & 63.89\% & 41.97\% & 50.66\% \\
			Measure\_Mods & 72 & 0.00\% & 0.00\% & 0.00\% & 33.33\% & 33.33\% & 33.33\% \\
			Network\_1 & 56 & 0.00\% & 0.00\% & 0.00\% & 70.69\% & 73.21\% & 71.93\% \\
			Network\_2 & 70 & 0.00\% & 0.00\% & 0.00\% & 53.85\% & 30.00\% & 38.53\% \\
			Pulse\_Gen & 95 & 0.00\% & 0.00\% & 0.00\% & 38.10\% & 42.11\% & 40.00\% \\
			Sensors & 40 & 0.00\% & 0.00\% & 0.00\% & 27.55\% & 67.50\% & 39.13\% \\
			Signal\_Gen & 56 & 0.00\% & 0.00\% & 0.00\% & 29.07\% & 44.64\% & 35.21\% \\
			Signal\_Proc & 167 & 0.00\% & 0.00\% & 0.00\% & 41.90\% & 26.35\% & 32.35\% \\
			Str\_Func & 177 & 0.00\% & 0.00\% & 0.00\% & 53.42\% & 48.59\% & 50.89\% \\
			Time\_and\_Date & 314 & 26.24\% & 69.43\% & 23.14\% & 66.90\% & 60.51\% & 63.55\% \\
			Vector\_Math & 57 & 0.00\% & 0.00\% & 0.00\% & 53.61\% & 91.23\% & 67.53\% \\
			\midrule
			Avg. & 2,430 & 5.11\% & 11.85\% & 4.90\% & 51.43\% & 49.38\% & 49.18\% \\
			\bottomrule
		\end{tabular*}
	\end{table*}
	
	To further inspect platform-level variation, we aggregate the ten-fold predictions of the hierarchical ACFG-GNN by platform, as shown in Table~\ref{tab:function-per-platform}. The platform-wise results show that CODESYS v3 has the lowest F1 (34.43\%), while GEB reaches 52.46\% F1 and OpenPLC v2 and v3 reach 46.67\% and 47.49\% F1, respectively. This gap suggests that our CODESYS-specific reverse-engineering tool is less effective compared to IDA, which we will further discuss in Section \ref{sec:codesysRE}.
	
	\begin{table*}[t]
		\centering
		\footnotesize
		\begin{minipage}[t]{0.41\textwidth}
			\centering
			\caption{Platform-wise functionality prediction results.}
			\label{tab:function-per-platform}
			\setlength{\tabcolsep}{3pt}
			\begin{tabular}{@{}lrrrr@{}}
				\toprule
				Platform & Number & Precision & Recall & F1 \\
				\midrule
				CODESYS v3 & 554 & 33.62\% & 37.91\% & 34.43\% \\
				GEB & 617 & 52.52\% & 55.21\% & 52.46\% \\
				OpenPLC v2 & 619 & 50.80\% & 49.79\% & 46.67\% \\
				OpenPLC v3 & 640 & 51.43\% & 50.61\% & 47.49\% \\
				\bottomrule
			\end{tabular}
		\end{minipage}
		\hfill
		\begin{minipage}[t]{0.55\textwidth}
			\centering
			\caption{Function-type ablation results.}
			\label{tab:ablation}
			\setlength{\tabcolsep}{3pt}
			\begin{tabular}{@{}llrrrr@{}}
				\toprule
				Task & Function set & Supp. & Prec. & Rec. & F1 \\
				\midrule
				\multirow{2}{*}{Toolchain} & Runtime (main) & 2,431 & 100.00\% & 100.00\% & 100.00\% \\
				& Core (abl.) & 2,431 & 74.19\% & 74.13\% & 74.16\% \\
				\cmidrule{1-6}
				\multirow{3}{*}{Functionality} & Core (main) & 2,430 & 51.43\% & 49.38\% & 49.18\% \\
				& Runtime (abl.) & 2,430 & 16.13\% & 7.78\% & 7.40\% \\
				& Core+runtime (abl.) & 2,430 & 48.50\% & 47.53\% & 47.38\% \\
				\bottomrule
			\end{tabular}
		\end{minipage}
	\end{table*}

	\noindent\textbf{Answering RQ~\ref{RQ-functionality}:}
	PLC-BinX reaches 51.43\% precision, 49.38\% recall, and 49.18\% F1 for functionality prediction over 22 labels. The result shows that recovering core functions and constructing function-level semantic representations facilitate the analysis of PLC binary functionality.
	
	\subsection{RQ~\ref{RQ-ablation}: Ablation Study on Function-Type Selection}
	
	RQ~\ref{RQ-toolchain} and RQ~\ref{RQ-functionality} separately use runtime functions and core functions for toolchain prediction and functionality prediction. To evaluate whether this selection is necessary, we design an ablation study that changes the function category used by each downstream task while reusing the same model architecture.
	
	For toolchain prediction, the main input runtime functions are replaced with core functions. This ablation evaluates whether core functions can substitute for runtime function features. 	
	For functionality prediction, we evaluate two variants: replacing core functions with runtime functions, and combining core functions with runtime functions. This complementary setting tests whether runtime functions can provide useful behavioral information for functionality prediction.
	
	Table~\ref{tab:ablation} reports the results of the corresponding function-type ablation settings. For toolchain prediction, the main runtime-function setting reaches 100.00\% precision, recall, and F1. Replacing runtime functions with core functions reduces the result to 74.19\% precision, 74.13\% recall, and 74.16\% F1. This result indicates that core functions cannot provide useful toolchain provenance information as runtime functions do. 	
	For functionality prediction, the main core-function setting reaches 51.43\% precision, 49.38\% recall, and 49.18\% F1. Runtime-only ACFGs reach 16.13\% precision, 7.78\% recall, and 7.40\% F1, indicating that runtime functions provide limited control behavior semantics. Core functions combined with runtime functions reach 48.50\% precision, 47.53\% recall, and 47.38\% F1, which is lower than the core-only setting. This suggests that adding runtime functions can obscure the control logic semantics needed for functionality prediction.
	
	\noindent\textbf{Answering RQ~\ref{RQ-ablation}:}
	Using core functions for toolchain prediction reduces F1 from 100.00\% to 74.16\%, and replacing core functions with runtime functions for functionality prediction reduces F1 from 49.18\% to 7.40\%.
	The ablation results support the task-specific function selection of PLC-BinX. Runtime functions provide more stable toolchain provenance features for toolchain prediction, while core functions provide the primary control logic semantics for functionality prediction.

	\section{Discussion}
	
	\subsection{Correctness of the CODESYS-specific reverse-engineering tool}
	\label{sec:codesysRE}
	
	Validating CODESYS v3 reverse engineering is difficult because CODESYS is a closed-source platform. Unlike an open compiler toolchain, it does not allow us to instrument the compiler during code generation, emit complete metadata about function boundaries, or generate a source-to-binary mapping that can serve as an oracle for the PLC-BEAD binaries. Therefore, we do not claim complete ground-truth reconstruction of all CODESYS functions. Instead, we validate the CODESYS reverse-engineering tool through two complementary validation sources.
	
	First, we audit a stratified sample of 100 CODESYS \texttt{.app} files. For each audited sample, after recovering CODESYS functions, we compare source-level function semantics with recovered binary core functions. The audit checks whether the semantics of recovered binary core functions, such as \texttt{PLC\_PRG}, are mapped to the semantics of source-level core functions. Among the 100 audited samples, 71 achieve full coverage of core functions, meaning that all core functions are recovered as independent functions. The remaining 29 samples have partial coverage of core functions. Across these samples, 272 of 317 source-level core function mappings are recovered, giving an audited mapping coverage of 85.80\%. The partial coverage cases still recover the main control logic, but may miss small functions such as bit-extraction functions, numeric-conversion functions, date-conversion functions, or setup functions. Some missing functions may be inlined, merged into caller functions, or transformed by the CODESYS compilation process. However, the closed-source toolchain prevents us from confirming this for every case.

	We further present several representative cases to illustrate their correspondence. In \texttt{CMP.app}, the recovered functions \texttt{PLC\_PRG}, \texttt{CMP}, \texttt{EXP10}, and \texttt{FLOOR} match the source program entry and its floating-point comparison helper chain. In \texttt{DT2\_TO\_SDT.app}, the recovered functions include \texttt{PLC\_PRG}, \texttt{DT2\_TO\_SDT}, \texttt{DAY\_OF\_MONTH}, \texttt{DAY\_OF\_WEEK}, \texttt{DAY\_OF\_YEAR}, \texttt{LEAP\_OF\_DATE}, \texttt{MONTH\_OF\_DATE}, and \texttt{YEAR\_OF\_DATE}, matching the source-level date and time decomposition logic. A partial-coverage example is \texttt{STAIR.app}: the recovered functions \texttt{PLC\_PRG} and \texttt{STAIR} match the source program entry and stair-step calculation logic, while the source-level numeric-conversion function \texttt{\_REAL\_TO\_DINT} does not appear as an independent recovered function. These observations show that the recovered CODESYS functions preserve meaningful source-level function structure in the inspected cases, while also exposing the mismatches in the recovery of small functions.
	
	Second, we use downstream-task performance as indirect evidence for CODESYS reverse-engineering quality. The recovered CODESYS functions and representations produced by PLC-BinX are converted into the same function-level semantic representations as GEB and OpenPLC, and are then used in both toolchain prediction and functionality prediction. The results reported in RQ~\ref{RQ-toolchain} and RQ~\ref{RQ-functionality} show that the recovered representations support toolchain prediction and functionality prediction, indicating that the CODESYS reverse-engineering tool preserves task-relevant function structure and semantic information for the downstream tasks. This result complements the manual inspection by showing that the recovered function-level semantic representations remain effective in downstream tasks of PLC binary analysis.
	
	Nevertheless, the functionality-prediction result on CODESYS is lower than those on GEB and OpenPLC. This gap indicates that our CODESYS-specific reverse-engineering tool provides meaningful recovery but is less effective when compared to the IDA-based analysis for standard binaries. In the future, we will continue to improve the effectiveness of the CODESYS-specific reverse-engineering tool and recover binary functions more precisely.
	
	\subsection{Failure Modes of PLC-BinX on Functionality Prediction}
	\label{sec:failuremode}
	
	Functionality prediction remains substantially harder than toolchain prediction. PLC-BinX obtains lower F1 in several difficult classes including \texttt{Arithmetic\_Func}, \texttt{Basic\_Other\_Func}, \texttt{Pulse\_Gen}, and \texttt{Latches\_FlipFlop\_ShiftReg}. One reason is label heterogeneity. For example, \texttt{Basic\_Other\_Func} represents the set of functions that do not fall into other functionality categories. As a consequence, its samples do not share unified function semantics and are difficult to classify.
	
	The imbalanced distribution of labels is another factor that affects functionality prediction. As shown in Table~\ref{tab:function-per-label}, several functionality categories contain relatively few binaries even after aggregating the ten folds, such as \texttt{Sensors} with 40 samples, \texttt{Arithmetic\_Func} with 44 samples, \texttt{Network\_1} and \texttt{Signal\_Gen} with 56 samples, and \texttt{Vector\_Math} with 57 samples. With fewer training instances in each fold, the model observes fewer binary-level variants of these functionalities, making the learned decision boundaries less stable.
	
	The performance differences across platforms also suggest that reverse-engineering quality affects downstream-task learning. CODESYS v3 has fewer named symbols and larger recovered functions, which can make it harder to conduct fine-grained semantic analysis. GEB and OpenPLC expose more conventional executable structures and generated names, which provide more explicit information for functionality prediction.

	\subsection{Use of Binary-Visible Symbols}
	
	PLC-BinX uses symbol information that can be recovered from the analyzed binary artifacts. These recovered symbols are used in two parts of the workflow. First, they provide binary-visible anchors for core function identification. For example, symbols or naming patterns such as \texttt{PLC\_PRG}, \texttt{dt\_PR\_*}, and \texttt{PROGRAM0\_body\_\_} help locate functions that implement the main PLC control logic. Second, when recovered names appear as function names or call targets, PLC-BinX keeps them as name-aware tokens in function-level semantic representations, such as \texttt{call\_name:\_ARRAY\_ABS}. In both cases, the symbols are extracted from the binary artifacts themselves and are not obtained from source code or manual annotations.
	
	This setting is consistent with the artifact-level analysis setting of PLC-BEAD and the prior PLC binary reverse-engineering work ICSREF. PLC-BinX uses the same PLC-BEAD binaries as the baseline dataset. Similarly, ICSREF also recovers available symbol and metadata information during CODESYS v2 binary reverse engineering. PLC-BinX follows this binary-only setting, but further uses the recovered symbols to support core function identification and function-level semantic representation.
	
	Recovered symbols are useful but not sufficient for PLC binary understanding. In practice, many recovered functions do not have corresponding symbols. Therefore, PLC-BinX does not represent PLC binaries only by recovered names. Its semantic representation is mainly built from two function-level views: normalized instruction sequences and ACFG structures. Recovered symbols are used as auxiliary binary-visible clues when available.

	\subsection{Scope of Downstream Tasks in PLC Binary Analysis}
	
	The goal of PLC-BinX is to provide a BCA framework to support PLC binary analysis by transforming heterogeneous compiled PLC binaries into function-level semantic representations. In this broader goal, toolchain prediction and functionality prediction serve as two representative downstream tasks for validation. They capture complementary aspects of PLC binary analysis: toolchain prediction evaluates whether recovered functions can reveal the platform or toolchain family, while functionality prediction evaluates whether recovered function semantics can support control-behavior recognition.
	
	The results on these two tasks suggest that PLC-BinX can provide task-relevant information beyond syntactic disassembly and can support cross-platform learning. Nevertheless, the scope of this evaluation remains limited. PLC binary analysis can also include vulnerability discovery, forensic auditing, control-flow and data-flow inspection, and safety-property verification. Therefore, our evaluation results can be viewed as evidence that PLC-BinX provides useful function-level semantic representations, rather than complete coverage of all PLC binary-analysis goals.
	
	\section{Threats to Validity}
	
	\textbf{Internal validity.}
	Reverse-engineering tools may introduce errors in function boundary recovery, function call recovery, or core function identification, and such errors can affect both the function-level semantic representations and the downstream-task prediction results. To reduce this risk, we use IDA Pro, one of the most mature and widely adopted disassembly tools, to implement IDA-based analysis for GEB and OpenPLC. For CODESYS, where IDA-based recovery is not directly applicable, we build a CODESYS-specific reverse-engineering tool and validate its recovered functions and representations through the manual inspection and downstream tasks. Nevertheless, some errors in recovered function boundaries, call targets, or the separation between control logic and runtime code may remain.

	\textbf{External validity.}
	Our evaluation uses PLC-BEAD binaries from CODESYS v3, GEB, OpenPLC v2, and OpenPLC v3. These platforms cover multiple binary forms, including \texttt{.app} containers and ARM ELF executables, but they do not cover all PLC vendors, processor architectures, compiler toolchains, or deployment configurations. Results may differ for platforms or compiler toolchains not represented in the dataset.
	
	\textbf{Construct validity.}
	The main construct threat comes from how the evaluation dataset and labels are created. Functionality labels are inherited from the PLC-BEAD source-program categories. Because our evaluation reuses PLC-BEAD, any errors in its dataset would be inherited by this study and could affect the measured performance.

	\section{Related Work}
	
	\subsection{PLC Binary Reverse Engineering}
	
	\textbf{General-purpose binary reverse engineering.} General-purpose reverse-engineering tools provide the foundation for analyzing conventional executable formats. IDA Pro~\cite{ida}, Ghidra~\cite{ghidra}, and Capstone~\cite{capstone} support disassembly and control-flow recovery across mainstream architectures, while angr~\cite{angr} provides a program-analysis platform for binary-level reasoning. Prior studies also show that accurate disassembly remains difficult even for standard binaries because code discovery, function boundary recovery, and indirect control flow recovery are error-prone~\cite{andriesse2016disassembly}. Recent learning-based work such as XDA further improves robust disassembly through transfer learning~\cite{xda}. These tools are suitable for PLC binaries that use standard executable formats, so PLC-BinX uses IDA-based disassembly and control-flow recovery for GEB and OpenPLC binaries. However, cross-platform PLC binary reverse engineering also requires additional PLC-specific processing. In particular, PLC-BinX must recover executable code from CODESYS \texttt{.app} containers and separate recovered functions into runtime functions for toolchain prediction and core functions for functionality prediction.
	
	\textbf{PLC binary reverse engineering.} PLC binaries have recently attracted increasing attention as they contain compiled logic that directly controls physical processes. Denial-of-engineering-operations attacks introduced Laddis to recover Allen-Bradley ladder logic from binaries~\cite{senthivel2018deo}, while SIMILO reconstructs PLC control logic for forensics by combining data-flow, control-flow, and syntax-flow information~\cite{qasim2019similo}. ICSREF provides an automated framework for reverse engineering CODESYS v2-compiled industrial control binaries~\cite{icsref}. CLEVER decompiles PLC applications into control logic~\cite{clever}, and CLADF combines reverse engineering and verification to detect and investigate control-logic attacks~\cite{geng2024cladf}. Benkraouda et al. further investigate PLC applications compiled by CODESYS 2.x and 3.x and highlight the need for PLC-specific binary-analysis tools~\cite{benkraouda2023plcspecific}. These studies demonstrate the feasibility of recovering PLC semantics from binaries. However, existing PLC reverse-engineering studies mainly focus on a specific PLC platform. Moreover, their outputs are often assembly code or decompiled logic, rather than function-level semantic representations that can support fine-grained cross-platform downstream tasks. In contrast, PLC-BinX targets a cross-platform PLC-binary setting and disassembles CODESYS v3 \texttt{.app} containers, GEB binaries, OpenPLC v2 binaries, and OpenPLC v3 binaries into function-level semantic representations.
	
	\subsection{PLC Binary Datasets and Representation Learning}
	
	\textbf{PLC binary datasets.} PLC-BEAD bridges the PLC binary analysis gap by providing a cross-compiler dataset of PLC binaries across multiple platforms and compiler settings~\cite{plcbead}. This dataset enables systematic evaluation beyond a single vendor, and it is the basis of our study. PLC-BinX reuses PLC-BEAD, but focuses on whether reverse engineering can construct function-level semantic representations from the compiled binaries rather than treating each binary only as a byte sequence.
	
	\textbf{PLC binary representation learning.} PLCEmbed introduces a neural framework for PLC binaries based on raw bytes~\cite{plcbead}. This design is valuable because it avoids directly handling heterogeneous binary formats. However, raw bytes provide limited access to higher-level semantic information such as functions, calls, control-flow structure, and recovered symbols. In contrast, PLC-BinX proposes a representation based on reverse engineering: it separates runtime and core functions, builds function-level semantic representations, and evaluates whether these representations support effective downstream tasks.
	
	\textbf{Traditional binary representation learning.} Binary-code similarity and representation learning have been extensively studied for standard binaries~\cite{haq2021survey,marcelli2022bfs}. Early cross-architecture bug-search systems such as discovRE and Genius rely on numeric features, CFG structure, and graph-based indexing to compare binary functions at scale~\cite{eschweiler2016discovre,feng2016genius}. Later learning-based methods move from handcrafted features to learned representations: Gemini learns graph embeddings from attributed CFGs~\cite{xu2017gemini}; VulSeeker and Order Matters enrich graph representations with semantic block features and order-aware neural modeling~\cite{gao2018vulseeker,yu2020ordermatters}; and DeepBinDiff learns program-wide representations for binary diffing~\cite{duan2020deepbindiff}. Sequence- and language-model-based approaches model instruction streams or execution semantics, including SAFE's self-attentive function embeddings~\cite{massarelli2019safe}, Asm2Vec's instruction-level representation for clone search~\cite{ding2019asm2vec}, PalmTree's assembly-language pretraining~\cite{li2021palmtree}, jTrans's jump-aware Transformer~\cite{wang2022jtrans}, and Trex's micro-trace-based execution semantics~\cite{pei2020trex}. Recent studies also show that compiler transformations such as function inlining complicate the assumed one-to-one function-matching granularity~\cite{jia2023inlining,jia2024crossinlining}. These works motivate our use of instruction tokens, ACFGs, and hierarchical aggregation, while PLC-BinX adapts these ideas to heterogeneous PLC binaries.
	
	\subsection{Related Research on PLC Security}
	
	\textbf{PLC testing.} Testing and fuzzing are another important line of PLC security research. ICSFuzz manipulates input and output values and repurposes binary code to enable instrumented fuzzing of ICS control applications~\cite{icsfuzz}. FieldFuzz performs in-situ black-box fuzzing of proprietary industrial automation runtimes through network interfaces~\cite{fieldfuzz}. ICSQuartz further introduces scan-cycle-aware fuzzing to align test injection with PLC execution semantics~\cite{icsquartz}. These approaches are effective for exposing runtime failures or vulnerable behaviors through execution. 
	
	\textbf{PLC formal verification.} Formal methods provide rigorous reasoning about PLC control logic and safety properties. PLCverif supports formal verification of PLC programs~\cite{plcverif}, and VoICS extends this direction to binary-level formal verification for PLC control logic in industrial IoT systems~\cite{zhang2025voics}. Such techniques motivate the auditing of deployed PLC binaries rather than source-level program checks. PLC-BinX shares this binary-level motivation, but focuses on reverse engineering and semantic representation for learning-based PLC binary analysis.

	\section{Conclusion}
	
	This paper presents PLC-BinX, a cross-platform BCA framework for PLC binaries. PLC-BinX applies a three-stage PLC binary analysis workflow: cross-platform reverse engineering, core function identification, and function-level semantic representation. We validate the recovered function-level semantic representations through two downstream tasks. On PLC-BEAD, PLC-BinX achieves 100.00\% precision, recall, and F1 for toolchain prediction, and 51.43\% precision, 49.38\% recall, and 49.18\% F1 for functionality prediction. These results demonstrate that PLC-BinX can expose task-relevant semantic information from heterogeneous PLC binaries and support effective PLC binary analysis.
	
	We hope our study could provide insights for subsequent studies and inspire more researchers to investigate the security analysis of PLC binaries.

	\bibliographystyle{ACM-Reference-Format}
	\bibliography{refs}

\end{document}